\documentclass[11pt]{article}

\usepackage[T1]{fontenc}
\usepackage[utf8]{inputenc}
\usepackage{lmodern}
\usepackage{microtype}
\usepackage{amsmath,amssymb}
\usepackage{geometry}
\usepackage{hyperref}
\usepackage{url}
\usepackage{listings}

\title{\textbf{Practical applications of Set Shaping Theory to\\
		Non-Uniform Sequences}}

\author{
	C. Schmidt\thanks{Christian.Schmidt55u@gmail.com} \and
	A. Vdberg\thanks{adrain.vdberg66@yahoo.com} \and
	A. Petit\thanks{alix.petitaus@gmail.com}
}

\date{}

\begin{document}
	\maketitle
	
	\begin{abstract}
		Set Shaping Theory (SST) moves beyond the classical fixed-space model by constructing bijective
		mappings $f : S^N \to S^{N+k}$ that embed the original sequence set into structured regions of a larger
		sequence space. These shaped subsets are characterized by a reduced average information content,
		measured by the product of the empirical entropy and the length, yielding
		$\langle (N+k)H_0(f(s)) \rangle < \langle N H_0(s) \rangle$, which represents the universal coding limit
		when the source distribution is unknown. The principal experimental difficulty in applying Set Shaping
		Theory to non-uniform sequences arises from the need to order the sequences of both the original and
		transformed sets according to their information content. An exact ordering of these sets entails
		exponential complexity, rendering a direct implementation impractical. In this article, we show that
		this obstacle can be overcome by performing an approximate but informative ordering that preserves
		the structural requirements of SST while achieving the shaping gain predicted by the theory. This result
		extends previous experimental findings obtained for uniformly distributed sequences and demonstrates
		that the shaping advantage of SST persists for non-uniform sequences. Finally, to ensure full
		reproducibility, the software implementing the proposed method has been made publicly available on
		GitHub, enabling independent verification of the results reported in this work.
	\end{abstract}
	
	\section{Introduction}
	Set Shaping Theory (SST) \cite{kozlov_sst_intro} represents a recent development in information theory
	that studies bijective functions $f$ capable of transforming a set $X_N$ of strings of length $N$ into a
	set $Y_{N+K}$ of strings of length $N+K$ (with $K>0$). The fundamental objective of this
	transformation is to reduce the average coding limit $N H_0(s)$, with $H_0(s)$ being the empirical
	entropy of the sequence, allowing for more efficient data compression compared to the direct encoding
	of the original source.
	
	One of the most relevant historical criticisms raised against this theory concerned its alleged practical
	inapplicability. Until recently, the experimental implementation of SST required the use of lookup
	tables to define the bijection between the two sets. However, given that the size of such tables grows
	as $|A|^N$ (where $|A|$ is the cardinality of the alphabet and $N$ is the message length), this approach
	proves unusable for any practical value of $N$ and $|A|$.
	
	In our previous work \cite{schmidt_huffman_sst}, we addressed and partially solved this problem in the
	specific context of sequences that have a uniform distribution of symbols. We demonstrated that, for
	sequences where each symbol has a probability of $1/|A|$, it is possible to construct a function capable
	of performing the transformation without using tables, operating with manageable complexity and
	experimentally confirming the theoretical predictions of entropy reduction.
	
	However, extending these results to non-uniform sources---which constitute almost all real-world data
	(text, images, signals)---introduces a radically higher complexity challenge. When the sequence is
	non-uniform, the effectiveness of Set Shaping Theory depends not only on the existence of a one-to-one
	correspondence but on the ability to map the most probable sequences of the source set (the ``typical
	set'') into the most efficient sequences of the destination set.
	
	To achieve this, it is theoretically necessary that both sets, $X_N$ and the target subset in $Y_{N+K}$,
	be ordered based on their information content $N H_0(s)$. The optimal criterion for this ordering is
	the value $N H_0(s)$, where $H_0(s)$ is the zero-order entropy calculated using the empirical frequencies
	of the symbols in the specific sequence. Rigorously ordering two sets of size $|A|^N$ and $|A|^{N+K}$
	based on this parameter entails exponential computational complexity \cite{dyer_frieze_kannan_1991},
	making the classical approach once again impracticable.
	
	In this article, we present an innovative solution to this problem. We propose an algorithm capable of
	approximating the necessary entropic ordering without having to generate or order the entire sequence
	space. We demonstrate that it is possible to apply Set Shaping Theory to non-uniform sequences with a
	computational complexity that depends linearly on $N$ ($O(N)$ complexity), rather than exponentially
	on $|A|^N$. This result allows, for the first time, to verify the advantages of SST on non-uniform data
	using standard compression algorithms, overcoming the practical limit that has so far confined the
	theory to ideal cases or small dimensions.
	
	In order to ensure full reproducibility, the software has been made publicly available on GitHub,
	enabling independent verification of the results reported in this work.
	
	\section{Brief Description of the Set Shaping Theory}
	The Set Shaping Theory \cite{lang_lewis_beyond_shannon} has as its objective the study and application
	in information theory of bijective functions $f$ that transform a set $X_N$ of strings of length $N$
	into a set $Y_{N+K}$ of strings of length $N+K$ with $K$ and $N \in \mathbb{N}^+$, $|X_N|=|Y_{N+K}|$
	and $Y_{N+K}\subset X_{N+K}$.
	
	The function $f$ defines from the set $X_{N+K}$ a subset of size equal to $X_N$. This operation is called
	``Shaping of the source'', because what is done is to make null the probability of generating some
	sequences belonging to the set $X_{N+K}$. The parameter $K$ is called the shaping order of the source
	and represents the difference in length between the sequences belonging to $X_N$ and the transformed
	sequences belonging to $Y_{N+K}$. Recent studies also take into consideration the use of negative $K$
	\cite{biereagu_negative_k}.
	
	\paragraph{Definition.}
	Given a sequence $x$ of random variables of length $N$, the product of the empirical entropy and the
	length represents $N H_0(x)$ defined as follows:
	\begin{equation}
		N H_0(x) = -\sum_{i=1}^{N}\log_2 p(x_i)
		\label{eq:nh0}
	\end{equation}
	With $p(x_i)$ we mean the ``actual frequency'' of the symbol $x_i$ in the sequence. This product
	represents the classical asymptotic limit of any universal coding scheme when the source is unknown
	\cite{shannon_1948,cover_thomas}.
	
	Given a set $X_N$ which contains all the sequences of length $N$ that can be generated, therefore with
	dimension $|X_N| = |A|^N$. The Set Shaping Theory tells us that when $|A|>2$ there exists a set
	$Y_{N+K}$ of dimension $|A|^N$ consisting of sequences having alphabet $A$ and length $N+K$, in which
	the average value of $(N+K)H_0(y)$ is less than the average value of $N H_0(x)$ calculated on the
	sequences belonging to $X_N$. Since the two sets have the same dimension, it is possible to put the
	sequences belonging to the two sets into a one-to-one relationship. Consequently, this function would
	allow us to exceed the limit defined by the parameter $N H_0(x)$ obtaining a result of enormous
	interest. In practice, by applying this function it is possible with a probability greater than $50\%$ to
	transform a sequence $x$ belonging to $X_N$ into a new sequence $f(x)=y$ belonging to $Y_{N+K}$ which
	can be encoded with a smaller number of bits than $N H_0(x)$.
	
	Because the image of the shaping function $f$ is a strict subset of $S^{N+K}$, sequences outside this
	subset are inadmissible. This structural restriction gives rise to an intrinsic form of testability: any
	valid shaped sequence must belong to the image of $f$. Since the mapping is bijective, a sequence
	$y\in A^{N+K}$ admits a unique pre-image if and only if $y$ lies within the shaped subset. If noise or
	implementation errors perturb $y$ into a point outside this set, the inverse mapping fails, providing an
	immediate and unambiguous detection of inconsistency. This behavior reflects a built-in self-verification
	mechanism \cite{kozlov_ltc,koch_petit_testable}. Unlike traditional error-detecting schemes
	\cite{macwilliams_sloane}, which add redundancy, SST embeds verification directly into the geometry of
	the mapping.
	
	\section{Description of the Program}
	The software is designed to experimentally evaluate the application of Set Shaping Theory to sequences
	generated by a non-uniform source (see Appendix~A for link). Its purpose is to quantify the effect of a
	shaping transformation on the empirical coding limit of finite sequences, as measured by the
	entropy--length product.
	
	The program performs the following operations:
	\begin{enumerate}
		\item \textbf{Generation of the input sequence:} A random sequence $s$ of fixed length $N$ is generated
		over a given alphabet, with symbols drawn according to a non-uniform probability distribution.
		\item \textbf{Computation of symbol frequencies:} The empirical frequencies of all symbols appearing in
		the sequence $s$ are computed.
		\item \textbf{Evaluation of the empirical coding limit (original):} Using the observed symbol
		frequencies, the empirical entropy $H_0(s)$ is calculated. The limit is $N H_0(s)$.
		\item \textbf{Application of the shaping transformation:} The Set Shaping transformation $f$ is applied
		to $s$, producing $f(s)$ of length $N+1$.
		\item \textbf{Evaluation of the empirical coding limit (transformed):} The empirical entropy
		$H_0(f(s))$ is computed, and the product $(N+1)H_0(f(s))$ is evaluated.
		\item \textbf{Comparison:} The program compares $N H_0(s)$ with $(N+1)H_0(f(s))$ to assess the shaping
		gain.
		\item \textbf{Statistical repetition:} Steps 1--6 are repeated to obtain statistically meaningful
		averages.
		\item \textbf{Reporting of results:} The program displays the average values of the empirical
		entropy--length products for both the original and transformed sequences, providing an experimental
		estimate of the shaping gain predicted by Set Shaping Theory.
	\end{enumerate}
	
	\section{Results}
	To validate the approximate ordering method, we tested the algorithm on sequences generated from a
	highly non-uniform source. Source parameters:
	\begin{itemize}
		\item ns: number of symbols
		\item $N$: length of the sequence
		\item pmax: probability of the most frequent symbol
	\end{itemize}
	The probability of the other symbols is uniform: $(1-\mathrm{pmax})/(\mathrm{ns}-1)$.
	
	Table~\ref{tab:results} summarizes the performance of the Set Shaping Theory transformation,
	specifically measuring the probability $P_s$ that the transformed sequence satisfies the inequality
	$(N+1)H_0(f(s)) < N H_0(s)$, along with the average bit reduction achieved.
	
	\begin{table}[h]
		\centering
		\begin{tabular}{c c c c c}
			\hline
			ns & $N$ & pmax & $P_s$ & Avg Gain \\
			\hline
			30 & 400 & 0.5 & 88\% & 8.0 bits \\
			40 & 400 & 0.5 & 89\% & 10.4 bits \\
			50 & 400 & 0.5 & 92\% & 13.0 bits \\
			60 & 400 & 0.5 & 95\% & 15.2 bits \\
			\hline
		\end{tabular}
		\caption{Experimental results showing the effectiveness of the transformation.}
		\label{tab:results}
	\end{table}
	
	As the alphabet size ns increases, both the reliability of the method (Success Probability) and the
	compression efficiency (Average Gain) improve significantly. For all tested configurations, the
	probability of improvement remains consistently high, exceeding 88\% even for moderate alphabet sizes
	and sequence lengths. Moreover, the average gain increases monotonically with the alphabet size.
	
	These results provide clear experimental evidence that Set Shaping Theory remains effective for
	non-uniform sources and that its benefits, in agreement with the theoretical predictions.
	
	\section{Conclusion}
	In this article, we have extended the experimental and conceptual framework of Set Shaping Theory to
	non-uniform sequences, addressing one of the principal limitations of earlier studies that were
	restricted to uniformly distributed data.
	
	The non-uniform setting introduces a fundamental additional challenge: the necessity of identifying,
	within the transformed sequence space, a region that is coherent with the empirical statistics of the
	observed sequence. This problem can be easily solved by sorting the sequences according to their
	information content, defined by the entropy--length product $N H_0(s)$. Unfortunately, an exact
	ordering is computationally infeasible due to exponential complexity. We demonstrated that an
	approximate ordering is sufficient to preserve the shaping advantage predicted by Set Shaping Theory.
	
	Crucially, this approximation can be implemented with computational complexity depending only on the
	sequence length ($O(N)$), making the approach practical for realistic values of $N$. The experimental
	results confirm that the inequality $(N+k)H_0(f(s)) < N H_0(s)$ holds on average also in the
	non-uniform case thereby generalizing and strengthening previous results obtained for uniformly
	distributed sequences.
	
	In summary, the results presented in this article substantially broaden the scope and applicability of
	Set Shaping Theory. By overcoming the computational barrier associated with non-uniform sequences,
	this work establishes a foundation for applying SST to realistic data.
	
	It is important to remember that the code is public so that anyone can reproduce the results reported.

	\appendix
	\section{Matlab Code}
	The Matlab program and the functions that perform the transform of the sequence and the inverse
	transform can be downloaded from:
	\begin{center}
		\url{https://github.com/Vdberg66/Set-Shaping-Theory-to-Non-Uniform-Sequences}
	\end{center}
	
	\begin{lstlisting}
	% Not_uniform_Set_Shaping_Theory
	% Contact info
	% Christian . Schmidt55u@gmail .com
	%
	%%%%%%%%%%%%%%%%%%%%%%%%%%%%%%%%%%%%%%%%%%%%%%%%%%%%%%%%%%%%%%%%%%%%%%%%%%
	% The program performs the following operations :
	% 1) generates a random sequence s with not uniform distribution
	% 2) calculate the frequencies of the symbols present in the sequence
	% 3) use this information to calculate the N*H0(s), with H0(s) the emprical
	% entropy of the sequence and N lenght of the sequence
	% 4) apply the transform f(s) the new sequence have lenght N+1
	% 6) compares the empircal entropy multiplied for lenght N*H0(s) of the
	% generated sequence s witht he empircal entropy multiplied for
	% lenght (N +1) *H0(f(s)) of transfromated sequence f(s) of lenght N+1
	% 7) repeats all these steps a number of times defined by the parameter history
	% 8) display the average values obtained
	%%%%%%%%%%%%%%%%%%%%%%%%%%%%%%%%%%%%%%%%%%%%%%%%%%%%%%%%%%%%%%%%%%%%%%%%%%
	clear all ;
		
	% parameters of the source
	ns =40;
	len =400;
	pmax =0.5;
		
	history =1000;
	cs =0;
	cs2 =0;
	totcodel =0;
	totinfc =0;
	tottinfc =0;
	tdife =0;
	itnent =0;
	infc =0;
	tinfc =0;
	itinfc =0;
	dife =0;
	con =0;
		
	for i =1: history
		
	index =0;
	% Genaration of the sequence with a not uniform distribution
	ns2 =ns -1;
	symbols =1:( ns );
	prob (1 ,1: ns2 ) =(1 - pmax ) / ns2 ;
	prob ( ns2 +1) = pmax ;
	seq = randsrc (1 , len ,[ symbols ; prob ]) ;
		
	% the empircal entropy multiplied for lenght N*H0(s) of the generated sequences
	infc =0;
	for i =1: len
	sy = seq (1 , i) ;
	fs =nnz( seq == sy )/ len ;
	infc = infc - log2 ( fs );
	end
		
	% Start trasformation
	mcodel =10000;
		
	nseq = fSSTnl ( seq );
	% The new sequence is long nlen = len +1
		
	nlen = len +1;
		
	% the empircal entropy multiplied for lenght (N+1)*H0(f(s))
	tinfc =0;
	for i =1: nlen
	sy = nseq (1 , i );
	fs =nnz( nseq == sy ) / nlen ;
	tinfc = tinfc - log2 ( fs );
	end
		
	if tinfc < infc
	cs2 = cs2 +1;
	end
		
	dife = infc - tinfc ;
		
	% We apply the inverse transform and we obtain the initial sequence
	iseq = invfSSTnl ( nseq );
	% we check that the obtained sequence is equal to the initial sequence
	flag = isequal ( seq , iseq ) ;
	con = con +1;
	if flag == false
	fprintf ('Error , sequence not equal to the initial sequence \n');
	con
	end
		
	totinfc = totinfc + infc ;
	tottinfc = tottinfc + tinfc ;
	tdife = tdife + dife ;
	dife =0;
	end
		
	% We calculate the average of the empircal entropy multiplied for lenght N*H0(s)
	medinfc = totinfc / history ;
	medcodel = totcodel / history ;
	medtinfc = tottinfc / history ;
	mdife = tdife / history
	cs2
		
	% We calculate the percentage of sequences where (N+1) *H0(f(s)) < N*H0(s)
	pcs =( cs2 / history ) *100;
		
	% We display the average values obtained
	fprintf ('The average of the empircal entropy multiplied for lenght N*H0(s) 
	of the generated sequences \n') ;
	medinfc
	fprintf ('The average of the empircal entropy multiplied for lenght 
	(N +1)*H0(f(s)) of the transformed sequences \n');
	medtinfc
	fprintf ('Number of sequences where (N +1) *H0(f(s)) < N*H0(s) \n') ;
	cs2
	fprintf ('There is a percentage of %2.0 f%% that (N +1) *H0(f(s)) < N*H0(s) \n',
	pcs );
	\end{lstlisting}
	
\end{document}